\documentclass[english,prb,twocolumn,showpacs,preprintnumbers,amsmath,
amssymb]{revtex4}
\usepackage[T1]{fontenc}
\usepackage[latin1]{inputenc}
\usepackage{verbatim}
\usepackage{textcomp}
\usepackage{graphicx}




\usepackage{babel}

\begin{document}

\title{Coherent control of localization, entanglement, and state superpositions 
in a double quantum dot with two electrons}

\author{G.\ E.\ Murgida}

\author{D.\ A.\ Wisniacki}

\author{P.\ I.\ Tamborenea}

\affiliation{Department of Physics {}``J.\ J.\ Giambiagi``{}, University
of Buenos Aires, Ciudad Universitaria, Pab.\ I, C1428EHA Buenos Aires,
Argentina}

\date{\today}

\begin{abstract}
We have recently proposed a quantum control method based on the knowledge of 
the energy spectrum as a function of an external control parameter  [Phys.\ 
Rev.\ Lett.\ {\bf 99}, 036806 (2007)].
So far, our method has been applied to connect the ground state to target
states that were in all cases energy eigenstates.
In this paper we extend that method in order to obtain more general 
target states, working, for concreteness, with a system of two interacting 
electrons confined in semiconductor double quantum wells.
Namely, we have shown that the same basic method can be employed to obtain
localization, entanglement, and general superpositions of eigenstates 
of the system.
\end{abstract}
\maketitle

\section{Introduction }
\label{sec:Introduction}

Quantum control is an area of research of great current interest with 
a vast potential for applications in quantum information technologies.
\cite{dal,JMO}
The basic problem of quantum control consists of driving externally a
quantum system in order to take it from an initial state to a target 
final state.
Quantum control ideas and methods are traditionally widely applied in 
magnetic resonance \cite{sli,van-chu} and quantum chemistry, \cite{sha-bru}
and they are advancing rapidly in the area of solid state nanostructures.
\cite{bon-etal,pet-etal,ber,rud}


In a series of recent publications \cite{mur-wis-tam,wis-mur-tam, mur-wis-tam-08} 
we have presented a method of quantum control based on the knowledge of the 
energy spectrum as a function of a suitable single control parameter.
This method is useful provided that the transitions between
neighboring levels are well described by the Landau-Zener model.\cite{zener}
The latter condition has allowed us to successfully navigate the spectrum
with a combination of diabatic and adiabatic changes of the control parameter.
Other authors have also employed the navigation of the energy
spectrum as a coherent quantum control tool, especially in the
field of atomic and molecular systems controlled by lasers.
\cite{vit-etal,gue-etal,yat-etal,yat-etal2,san-etal}
We have applied our control method to two different physical
systems with remarkable success.
The first one, which we will study in this paper, is a nanostructured
semiconductor system consisting of two interacting electrons confined
in quasi-one-dimensional quantum dots,\cite{mur-wis-tam,wis-mur-tam,
mur-wis-tam-08} and the second one was the LiCN molecule.\cite{licn}
In both systems we have been able to connect distant eigenstates through
long and complex paths in the energy spectrum with very high probability.


In this paper we take the control method farther: by allowing not only 
diabatic and adiabatic transitions but also intermediate velocities, 
we can arrive at more complex final (target) states.
With this generalized navigation method we are able to achieve the creation
of linear superpositions of adiabatic states starting from the ground
state.
This opens a rich menu of possibilities, like the creation of Bell states
and other types of entangled states.

The paper is organized as follows: in Section \ref{sec:method} we describe
the main ingredients of our control method.
In Section \ref{sec:system} we present again the two-electron nanostructure.
Section \ref{sec:results} is devoted to the results of the generalized
control method. 
Among the results presented, we show how to apply our method to generate 
Bell states and coherent superpositions of energy eigenstates.

\begin{figure}
   \includegraphics[angle=-90,width=8.5cm]{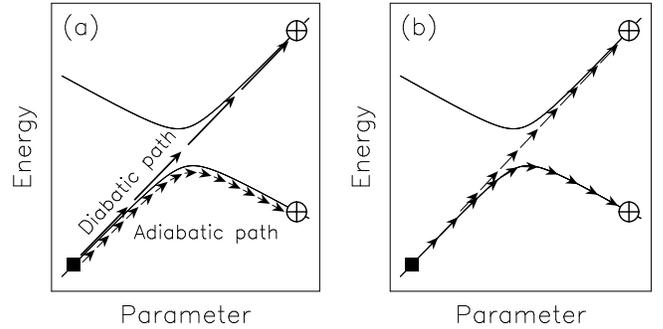}
   \caption{Schematic plot of the building block of our control method.
            An avoided crossing and a schematic drawing of the 
            diabatic (fast), adiabatic (slow), 
            and intermediate ways to cross it.
            (a) Long (short) arrows represent fast (slow) variations of 
            the control parameter;
            (b) Medium arrows represent the intermediate velocity.}
   \label{fig:LZ} 
\end{figure}

\section{The control method: Navigating the Hilbert space}
\label{sec:method}

Let us first review the basic ideas of our method in the simplest possible 
system, i.e.\ a parameter-dependent two level system.
Let us assume that these two levels have an avoided crossing as shown in 
Fig.\ \ref{fig:LZ}.
At the avoided crossing, the two energy levels approach each other and 
the associated eigenstates exchange their characteristics as the external 
parameter sweeps through the crossing.
If the avoided crossing is traversed very slowly (adiabatic path in 
Fig.\ \ref{fig:LZ}(a)), the adiabatic theorem guarantees that one will stay 
in the initial adiabatic level, but with the paradoxical consequence that 
the final state will have been exchanged with the other diabatic state. 
On the other hand, if the avoided crossing is traversed very quickly
(diabatic path in Fig.\ \ref{fig:LZ}), the characteristics of the state 
are preserved.
Clearly, these two limiting possibilities suggest a very simple control method.
The quantitative meaning of slow and fast transitions is given by
the theory of Landau-Zener transitions.\cite{zener, mur-wis-tam-08}

This basic idea has allowed us to perform complicated control tasks.
Namely, we can travel through the energy spectrum using the avoided
crossings as sources of efficient controllable choices between two adiabatic
states.
In this way we can connect distant adiabatic states provided that there exists
a path going from one to the other via jumps at avoided crossing and
adiabatic evolutions in the absence of crossings. 
Examples of the application of this method have been presented in 
Refs.\ [\onlinecite{mur-wis-tam,wis-mur-tam}].

In this paper we go one step further by using not only diabatic and
adiabatic transitions but also transitions with intermediate speeds.
This type of transition gives final states (on exit of the avoided
crossing) which are linear combinations of the two adiabatic states
(Fig.\ \ref{fig:LZ}(b)).
The combination of several crossings using intermediate velocities 
allows us to access a great variety of final states.
Thus, in this work we generalize the control method proposed earlier
and present several applications which show the power of the improved
control technique.


\section{The double quantum dot system}
\label{sec:system}

\begin{figure}
  \includegraphics[angle=-90,width=7.5cm]{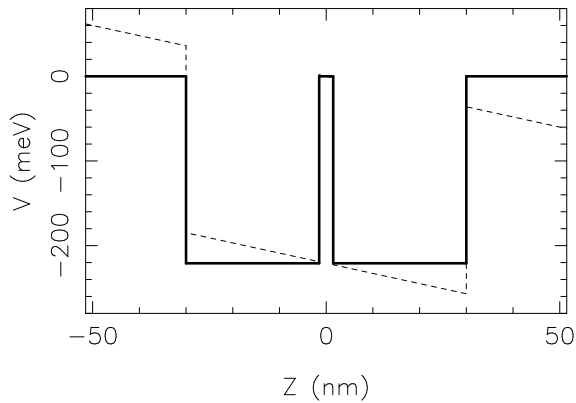}
  \caption{Confining double well potential in the longitudinal direction of the coupled quantum dot structure. The external electric field is  $E=0$ (solid lines) and $E=12 \, \mbox{kV/cm}$ (dashed lines).}
  \label{fig:double-dot}
\end{figure}
In order to explore on a concrete system the idea of using intermediate
speeds to traverse avoided crossings we will study here a system of
two interacting electrons inside a quasi-one-dimensional double quantum dot
which was used in our previous works.
The system is subject to an external, uniform electric field, whose 
amplitude is used as the control parameter.
The confinement of the two electrons is very narrow on two dimensions,
which we denote as $x$ and $y$, and the double well profile is defined
along the remaining, longitudinal coordinate, $z$.
The effective Hamiltonian of the two electrons reads
\begin{eqnarray}
H&\equiv&-\frac{\hbar^2}{2m}
      \left(\frac{\partial^2}{\partial z_1^2} + 
            \frac{\partial^2}{\partial z_2^2}
      \right) + V(z_1) + V(z_2) \nonumber
 \\
&&+ V_C(|z_1-z_2|) - e(z_1 + z_2)E(t),
\label{eq:hamiltonian}
\end{eqnarray}
where $m$ is the effective electron mass in the semiconductor material, 
$V_C$ is the effective Coulomb interaction between the electrons, 
\cite{tam-met-99} 
$V(z)$ is the confining potential (see Fig.\ \ref{fig:double-dot}), 
and $E(t)$ is the external time-dependent electric field.
We choose as confining potential a double quantum well with well width of 28~nm,
interwell barrier of 4~nm, and 220~meV deep (a typical depth for a GaAs-AlGaAs
quantum well).
We normally assume that the initial state is the ground state, which is 
a singlet.
Since the Hamiltonian is spin independent, the total spin is conserved 
and the spatial wave function remains symmetric under particle exchange
at all times.

\begin{figure}
\includegraphics[angle=-90,width=8.5cm]{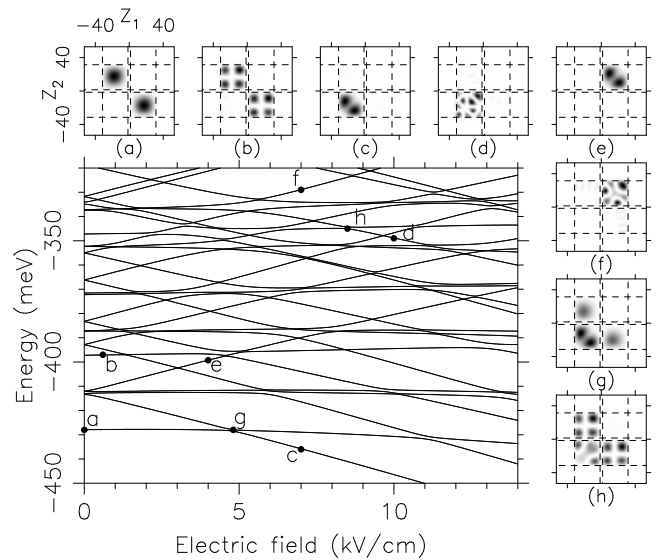}
\caption{
   {\it Main panel}: The energy spectrum of the two interacting electrons
    confined in a quasi-one-dimensional double-well semiconductor 
    nanostructure as a 
function of an external uniform electric field. 
The first 31 energy levels are shown. 
{\it Side panels}: 
Spatial wave functions $\phi_i(E,z_1,z_2)$ corresponding to labels 
(a) to (h) of the main panel.
States (a) and (b) have one electron in each well, wave functions 
(c) and (d) are localized in the left well, and wave functions (e) and (f)
have both electrons in the right well. 
States (a) to (f) are far from avoided crossings and therefore have 
well-defined localization properties. 
This is not the case for eigenstates (g) and (h), which are at avoided 
crossings.}
\label{fig:spectrum}
\end{figure}

Let us review some characteristics of the energy spectrum of the two electrons
and of the eigenstates which will guide our control strategies.
First we consider the spectrum as a function of the control parameter (the
external electric field).
This spectrum is plotted in Fig.\ \ref{fig:spectrum} with selected
eigenstates.
The energies $\epsilon_i(E)$ and the eigenstates $\phi_i(E,z_1,z_2)$
are obtained by numerical diagonalization.
For this calculation, we have used as basis set the symmetric combinations of 
the twelve lowest single-particle eigenfunctions. 
Thus, our basis set of the two-particle Hilbert space has 12*(12+1)/2=78 states.
An important characteristic of this system in the lower part of its spectrum
is that the states have fairly well-defined localization properties.
The three possible types of localization (both electrons in the left dot,
in the right dot, or electrons in different dots) are associated with 
the three types of slopes seen in the spectrum (positive, negative, or
almost zero, respectively).
In Fig.\ \ref{fig:spectrum} we show some states to illustrate the
mentioned localization characteristics.
In states a and b the electrons are in different dots,
in states c and d both electrons are in the left dot, and
in states e and f both electrons are in the right dot.   
Of course, at the avoided crossing the states get mixed and these
well-defined properties are lost (states (g) and (h)).


\section{Results}
  \label{sec:results}

This section is devoted to illustrate the power of the generalized control 
method in which the velocities to cross the avoided crossings are not
restricted to produce diabatic and adiabatic transitions.
In other words, we will show the possibilities opened by the ability
to allow intermediate velocities.
Due to the potential for applications of our method, we consider it best
to illustrate its power through two important examples of coherent
control.


\subsection{Localization}
  \label{sec:localization}

An application of quantum control that has been extensively explored in
the recent literature is the localization of one or two electrons in
a double well potential.
The general idea is to start the coherent evolution in the ground
state, which in the case of two electrons is delocalized due to
the Pauli and Coulomb repulsions, and end up in a state in which
both electrons are in the same well. 
In order to describe the degree of localization of the two electrons
we define the probability for both electrons to be in the left well,
$P_{\text{LL}}$.
\begin{equation}
P_{\rm{LL}}(t) = \int_{-\infty}^0 
                 \int_{-\infty}^0 dz_1 dz_2 \, |\psi(z_1,z_2,t)|^2, 
\end{equation}
where $\psi(z_1,z_2,t)$ is the evolving wave functions of the two electrons.

\begin{figure}
  \includegraphics[angle=-90,width=8.5cm]{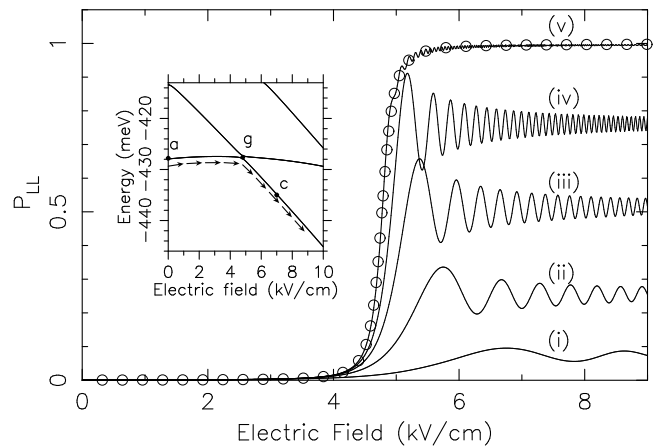}
  \caption{Probability $P_{\text{LL}}$ as a function of the time-dependent
           electric field $E(t)$.
           The velocities of the electric field are: (i) 4, (ii) 1, 
           (iii) 0.4, (iv) 0.2, and (v) 0.04 (kV/cm)/ps.
           With circles we represent $P_{\text{LL}}$ of the adiabatic
           ground state as a function of the electric field (see text for
           details.)
           Inset: closer view of the first avoided crossing involved in
           these processes. The arrows indicate the adiabatic path.}
  \label{fig:avoided_1}
\end{figure}

An easy way to localize both electrons in the left dot using our control
method starting from the ground state, consists of varying the electric
field adiabatically in order to pass the first avoided crossing between
the first two levels located at $E = 4.81 \, \mbox{kV/cm}$.
This process is shown in the inset of Fig.\ \ref{fig:avoided_1}.
(The numerical solution of the time-dependent Schr\"odinger equation 
was obtained using the usual fourth-order Runge-Kutta method with a 
time step of $0.05 \, \mbox{fs}$.)
The electric field varies linearly with time and we display the evolution
of $P_{\text{LL}}$ for several velocities. 
We start at the ground state, and accordingly $P_{\text{LL}}\simeq 0$ 
at $t=0$.
For a high velocity, we expect to cross diabatically the avoided
crossing and the localization properties will not change considerably.
This case is seen in curve (i) ($\dot{E}= 4 \, \mbox{(kV/cm)/ps}$.) 
in Fig.\ \ref{fig:avoided_1}.
As the velocity is decreased (curves (ii) to (v)) we approach a
final state that is highly localized in the left dot.
%
%
There is a maximum value of the probability $P_{\text{LL}}$ that can be
obtained with this method, given by the probability $P_{\text{LL}}$ 
of the adiabatic ground state after the avoided crossing.
The probability of the adiabatic ground state is plotted with 
open circles in Fig.\ \ref{fig:avoided_1}. 
After the crossing it is a slowly rising function in the plotted range 
and reaches the value of 0.998 at electric field $E = 9 \, \mbox{kV/cm}$.
We see that for the velocity of curve (v) 
($\dot{E} = 0.04 \, \mbox{(kV/cm)/ps}$) 
the evolving probability $P_{\text{LL}}(t)$ follows tighlty the open circles 
and near $E = 9 \, \mbox{kV/cm}$ presents a small oscillation bound  
between 0.995 and 0.9975.
That is, the maximum value of $0.998$ is approached within a $0.3$ percent.
%


\begin{figure}
  \includegraphics[angle=-90,width=7.5cm]{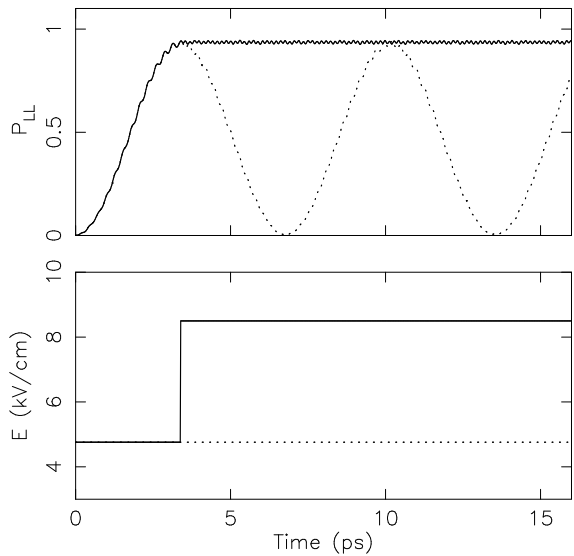}
  \caption{
           Localization with the sudden-switch method.
           While the electric field stays at the avoided-crossing value
           ($E = 4.81 \, \mbox{kV/cm}$) the probability $P_LL$ oscillates 
           with the frequency corresponding to the energy gap (dashed lines).
           The method consists of increasing suddenly the electric field 
           when $P_LL$ is maximal, freezing the localization on its highest
           value (full lines).
           Upper panel: Probability that both electrons are in the left dot.
           Lower panel: Step-wise constant electric field."
          }
  \label{fig:sudden-switch}
\end{figure}

Another simple way to obtain localization in this system  
is what we may call the sudden-switch method, which uses step-wise 
constant fields.\cite{tam-met-01} 
We wish to compare the effectiveness of both methods to obtain
localization.
First, let us summarize the basic procedure of the sudden-switch
method (See Fig.\ \ref{fig:sudden-switch}).
Starting from the ground state one applies two successive steps 
of constant electric field. 
In the first step one goes from zero field to the field corresponding 
to the first avoided crossing, i.e. $E=4.81 \, \mbox{kV/cm}$ in our case.
While the new field is on, the probability $P_{\rm{LL}}$ oscillates with the frequency corresponding to the energy splitting at the
avoided crossing.
This oscillation occurs because the initial state is no longer
eigenstate of the Hamiltonian at the avoided crossing.
In fact, it is a 50\%--50\% linear superposition of the two adiabatic 
states (eigenstates) at the avoided crossing.
When the probability $P_{\rm{LL}}$ reaches a maximum, that means that
the current state is the other diabatic state.
At this time one switches again the field to leave the avoided crossing.
Far from the avoided crossing the diabatic states are very close to the 
eigenstates of the Hamiltonian, and therefore, are approximately
stationary.
Thus, the $P_{\rm{LL}}$ remains at the highest value attained during
the oscillations at the avoided crossing.
With this method, a localization $P_{\rm{LL}}$ of up to 93\% can be
achieved, as can be seen in Fig.\ \ref{fig:sudden-switch}.
While the time scales involved in both methods are the same,
in comparison, our method has three advantages: 
(i) a higher degree of localization can be obtained; 
(ii) the sudden-switch method requires a fine adjustment of the timing 
not needed in our case;
(iii) our method is more powerful in the sense that can be used 
to navigate in the spectrum and connect distant states.\cite{mur-wis-tam} 

\begin{figure}
  \includegraphics[angle=-90,width=7.5cm]{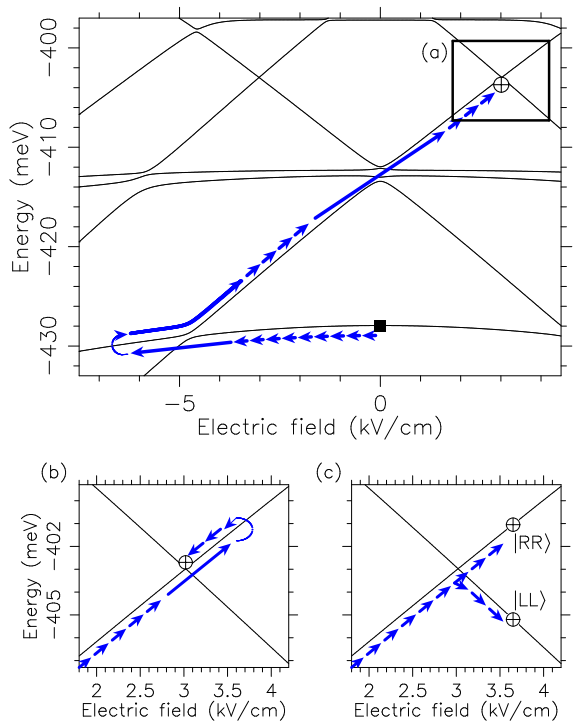}
  \caption{Control paths for two electrons leading to target states which are
           superpositions of states with RR and LL localization.
           The lengths of the arrows indicates the velocity of the transitions.
           The target states are:
           (a) $(|RR\rangle + |LL\rangle)/\sqrt{2}$;
           (b) $(|RR\rangle - |LL\rangle)/\sqrt{2}$;
           (c) state given in Eq.\ (\ref{eq:RRLL}).
          }
  \label{fig:RRLL}
\end{figure}

Let us now consider states with a different and in a sense more
complex kind of localization.
As mentioned earlier, we can find in the spectrum three types of localized
states, which we can denote in the following way: $|RR\rangle$, $|LL\rangle$,
and $|RL \rangle $, which have, respectively, both electrons in
the right and left dots, and one electron in each dot. 
(Of course, these states cannot be considered to be product states of 
single-particle orbitals, because quantum correlations are generally
present in all of them.\cite{cir-etal})
We usually refer to the first two types as localized states, and to the
third one as delocalized.
However, a linear superposition of the first two types can also be
considered as localized, in the sense that one knows that both electrons
would be found in the same dot if a measurement were performed.
In Fig.\ \ref{fig:RRLL} we show some control paths that may be followed in the spectrum
to reach a superposition of $|RR\rangle$ and $|LL\rangle$.
Linear superpositions of the form $(|RR\rangle \pm |LL\rangle)/\sqrt{2}$ 
are always available at the center of avoided crossings of states with 
RR and LL localization.
In Fig.\ \ref{fig:RRLL} (a) and (b) we show how to go from the ground state to the Bell-type states 
$(|RR\rangle + |LL\rangle)/\sqrt{2}$ and $(|RR\rangle - |LL\rangle)/\sqrt{2}$,
respectively, using only diabatic and adiabatic transitions.
We remark that these states are energy eigenstates, and thus do not evolve
further if the electric field is fixed.
If the restriction of using only diabatic and adiabatic transitions is relaxed
we can traverse the avoided crossing of the two types of states with an 
intermediate speed thus obtaining a combination of the diabatic states
after the crossing 
\begin{equation}
a \, e^{-i E_{RR} t / \hbar} |RR\rangle + 
b \, e^{-i E_{LL} t / \hbar} |LL\rangle.
\label{eq:RRLL}
\end{equation}
The values of $|a|$ and $|b|$ can be tuned by choosing the appropriate velocity.
We note that the effects of using intermediate velocities was illustrated 
previously (Fig.\ \ref{fig:avoided_1}) in the context of searching for a LL
localized state.

\begin{figure}
  \includegraphics[angle=-90,width=7.5cm]{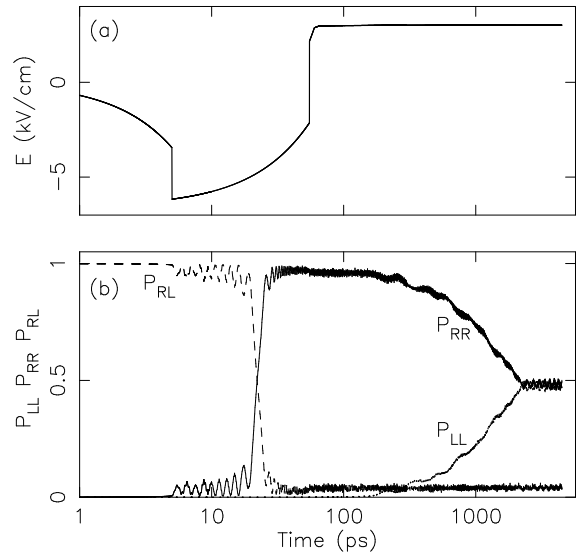}
  \caption{Numerical simulation of the path of Fig.\ \ref{fig:RRLL}(a) in order to reach the Bell-type state $(|RR\rangle + |LL\rangle)/\sqrt{2}$. The electric field used (a) and the probabilities obtained (b) are shown as functions of time in semi-log plots.
          }
  \label{fig:RRLLtime}
\end{figure}

We now show in Fig.\ \ref{fig:RRLLtime} how the path (a) of Fig.\ \ref{fig:RRLL} 
is obtained by varying the electric field appropriately.
In Fig.\ \ref{fig:RRLLtime} (a) we show the electric field and in (b) we show the probabilities
as a function of time (semi-log plot).
To arrive at the desired state, we must cross two avoided-crossing adiabatically,
one at $E= -4.81 \, \mbox{kV/cm}$ and energy $=-437 \, \mbox{meV}$, and the other one at
 $E= 3.02 \, \mbox{kV/cm}$ and energy $=-402 \, \mbox{meV}$.
The first avoided crossing involves a RL and a RR state, and the second one,
RR and LL states. 
The energy gap of these avoided crossings are very different, and therefore
the time required to cross them adiabatically is also very different.
This is way the log scale for the time axis is used in Fig.\ \ref{fig:RRLLtime}.
We see that once the electric field is left constant the probabilities $P_{RR}$ 
and $P_{LL}$ oscillate between 0.465 and 0.495 while the overlap of the final state with the target one es equal to 0.98, showing that we have arrived at the desired state.
The other two paths shown in Fig.\ \ref{fig:RRLL} can be realized in a similar
way, and with the same degree of success.

\subsection{Coherent superpositions}
\label{sec:superpositions}

In the previous section we used our method to control the localization properties
of the target state.
Now we illustrate the flexibility of our method by obtaining coherent superpositions
of several adiabatic eigenstates.
For example, we may wish to obtain a linear superposition of the three types of
localization having each of them the same weight.
That is, we seek a state of the form
\begin{equation}
|\psi \rangle = \frac{1}{\sqrt{3}} 
                \left( a |RR\rangle + b |LL\rangle + c |RL\rangle \right),
\label{eq:3superposition}
\end{equation}
with $|a|=|b|=|c|=1$.

\begin{figure}
   \includegraphics[angle=-90,width=7.5cm]{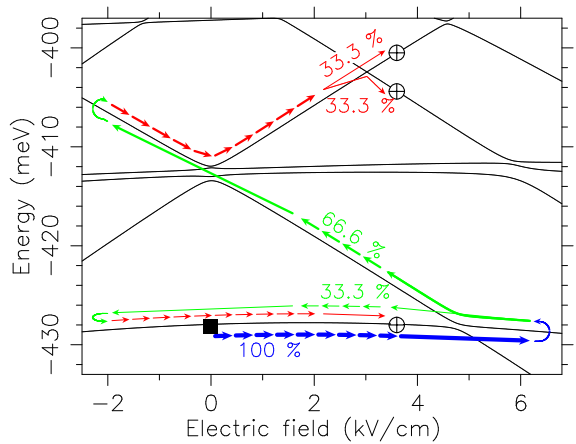}
   \caption{(Color online) Schematic diagram of the path to obtain the target which 
             is a superposition of three adiabatic states with different 
             localization (Eq.\ \ref{eq:3superposition}).
             For clarity, we use different colors when the electric field is increasing
             (blue and red) or decreasing (green).
             Our initial state, the ground state without electric field, is indicated 
             by a filled square).
             The desired target state is a combination of the three adiabatic states
             indicated by $\oplus$.}
   \label{fig:3superposition}
\end{figure}

\begin{figure}
   \includegraphics[angle=-90,width=7.5cm]{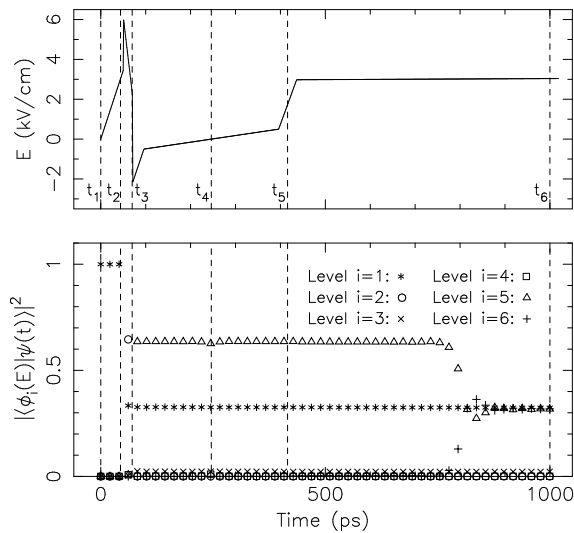}
   \caption{Upper Panel: Electric field as a function of time used to carry out the control
            strategy schematically shown in Fig.\ \ref{fig:3superposition}.
            Lower Panel: Overlap of the evolving state with the first six adiabatic states.  
            We can clearly see that the final state approximates well an even linear
            combination of the adiabatic states 1, 5, and 6 at the corresponding final
            electric field.}
   \label{fig:3sup_field_and_overlap}
\end{figure}

\begin{figure}[hbp]
   \includegraphics[angle=-90,width=7.5cm]{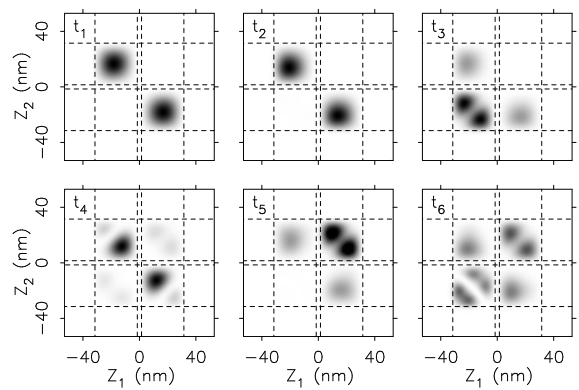}
   \caption{Evolving wave function at various times during the time evolution schematically
            shown in Fig.\ \ref{fig:3superposition} and described quantitatively in
            Fig.\ \ref{fig:3sup_field_and_overlap}.
            See text for details.}
   \label{fig:3sup_wavefunctions}
\end{figure}

In Fig.\ \ref{fig:3superposition} the control path to such a state is shown.
In Fig.\ \ref{fig:3sup_field_and_overlap} we show the overlap of the evolving 
state with the first six adiabatic states and the electric field (the control parameter) as
a function of time.
To appreciate the evolution in greater detail we plot the evolving
state at chosen times in Fig.\ \ref{fig:3sup_wavefunctions}.
The initial state is the ground state 
(see Fig.\ \ref{fig:3sup_wavefunctions}, state at time $t_1=0$), and the final
state is a superposition of the states 1, 5, and 6 at the electric field $E= 3.04 \, \mbox{kV/cm}$
(see Fig.\ \ref{fig:3sup_wavefunctions}, state at time $t_6 = 1000 \, \mbox{ps}$). 
We start the evolution by increasing slowly the electric field 
(blue arrows in Fig.\ \ref{fig:3superposition}), 
so that the state remains at the ground state 
(see Fig.\ \ref{fig:3sup_wavefunctions}, state at time $t_2$), 
then we accelerate to cross diabatically the first avoided crossing at 
$E = 4.81 \, \mbox{kV/cm}$, 
and then we decrease the field 
(green arrows in Fig.\ \ref{fig:3superposition})
in order to cross the same avoided crossing in the opposite direction 
with an intermediate velocity.
Here the occupation probability is split between the ground state (33.3\%)
and the first excited state (66.6\%).  
This is clearly seen at time $t_3$ in Fig.\ \ref{fig:3sup_field_and_overlap}. 
The electric field is further decreased, slowly at first and then rapidly, 
so that the upper branch crosses the complex avoided crossing at $E=0$
and energy of $-413 \, \mbox{meV}$.
Then an adiabatic increase follows.
We again cross the complex avoided crossing (time $t_4$ in the middle of the
avoided crossing and time $t_5$ after crossing it adiabatically).
We see that the state at $t_5$ is a combination of a RL and a RR state.
We continue slowly until the upper branch approaches the last avoided 
crossing at $E = 3.02 \, \mbox{kV/cm}$.
Here the velocity is chosen at an intermediate value, so that the occupation 
probability divides itself equally among the two states.
The end result is a superposition of the three states mentioned at
the beginning.
This is clearly verified in Figs.\ \ref{fig:3sup_field_and_overlap} and\ \ref{fig:3sup_wavefunctions}, state at $t_6 = 1000 \, \mbox{ps}$.
In this simulation, the final square overlap with the adiabatic states 1, 5, 
and 6 are respectively $P_1 = 0.324$, $P_5 = 0.325$ and $P_6 = 0.320$, and 
then the final probability to find the target state is of 96.45\%. 

\section{Final Remarks}

In this paper we extended a recently proposed control method for quantum systems
with energy spectra, which, as function of a control parameter, are characterized
by the presence of avoided crossings.  
While in previous publications we showed how to work with diabatic and adiabatic
changes of the control parameter, that is, using the avoided crossing as a
yes-no switch, here we explored the possibility of using intermediate
velocities to cross the avoided crossings, thus obtaining linear combinations of
diabatic states.
This generalization of our previous control strategy enables us to reach
more general target states, not restricted to eigenstates of the 
system's Hamiltonian.

The results presented here are for a two-electron double quantum-dot structure, 
but are not restricted to that particular system.
Since the issue of localization is an interesting one in double-well 
potentials, we have studied it here as an application of our method.
We showed how one can obtain target states with different types of 
localization, like having both electrons in one given well or having
both electron in either well with chosen probabilities (Bell-type states).
Finally, we have shown how to obtain a coherent superposition of three
states with each of the three types of localization present in this system.

\label{final_remarks}

\section*{Acknowledgements}
The authors acknowledge the support from CONICET (PIP-6137, PIP-5851) and
UBACyT (X237, X495).
D.A.W.\ and P.I.T.\ are researchers of CONICET.


\end{document}